\documentclass[prl,superscriptaddress,twocolumn,amsmath,amssymb,showpacs,nobibnotes]{revtex4-1}

\usepackage{graphicx}
\usepackage{txfonts}
\usepackage{bm}
\usepackage[dvips]{epsfig}
\makeatletter
\def\btt#1{\texttt{\@backslashchar#1}}%
\DeclareRobustCommand\bblash{\btt{\@backslashchar}}%
\makeatother

\begin{document}
\title{A Cooper pair light emitting diode}

\author{H.\ Sasakura}
\email[]{hirotaka@eng.hokudai.ac.jp}    
\affiliation{Research Institute for Electronic Science, Hokkaido University, Sapporo 001-0021, Japan}
\affiliation{CREST, Japan Science and Technology Agency, Kawaguchi 332-0012, Japan}

\author{S.\ Kuramitsu}
\affiliation{Research Institute for Electronic Science, Hokkaido University, Sapporo 001-0021, Japan}
\affiliation{Graduate School of Information Science Technology, Hokkaido University, Sapporo 060-0814, Japan}

\author{Y.\ Hayashi}
\affiliation{Research Institute for Electronic Science, Hokkaido University, Sapporo 001-0021, Japan}

\author{K.\ Tanaka}
\affiliation{CREST, Japan Science and Technology Agency, Kawaguchi 332-0012, Japan}
\affiliation{Central Research Laboratory, Hamamatsu Photonics, Hamamatsu 434-8601, Japan}

\author{T.\ Akazaki}
\affiliation{CREST, Japan Science and Technology Agency, Kawaguchi 332-0012, Japan}
\affiliation{NTT Basic Research Laboratory, Atsugi 243-0198, Japan}

\author{E.\ Hanamura}
\affiliation{Japan Science and Technology Agency, Kawaguchi 332-0012, Japan}

\author{R.\ Inoue}
\affiliation{CREST, Japan Science and Technology Agency, Kawaguchi 332-0012, Japan}
\affiliation{Department of Applied Physics, Tokyo University of Science, Tokyo 162-8601, Japan}

\author{H.\ Takayanagi}
\affiliation{CREST, Japan Science and Technology Agency, Kawaguchi 332-0012, Japan}
\affiliation{Department of Applied Physics, Tokyo University of Science, Tokyo 162-8601, Japan}

\author{Y.\ Asano}
\affiliation{Graduate School of Engineering, Hokkaido University, Sapporo 060-8628, Japan}

\author{H.\ Kumano}
\affiliation{Research Institute for Electronic Science, Hokkaido University, Sapporo 001-0021, Japan}
\affiliation{CREST, Japan Science and Technology Agency, Kawaguchi 332-0012, Japan}

\author{I.\ Suemune}
\email[]{isuemune@es.hokudai.ac.jp}   
\affiliation{Research Institute for Electronic Science, Hokkaido University, Sapporo 001-0021, Japan}
\affiliation{CREST, Japan Science and Technology Agency, Kawaguchi 332-0012, Japan}

\date{\today}

\begin{abstract}
We demonstrate Cooper-pair's drastic enhancement effect on band-to-band radiative recombination in a semiconductor. Electron Cooper pairs injected from a superconducting electrode into an active layer by the proximity effect recombine with holes injected from a p-type electrode and dramatically accelerate the photon generation rates of a light emitting diode in the optical-fiber communication band. Cooper pairs are the condensation of electrons at a spin-singlet quantum state and this condensation leads to the observed enhancement of the electric-dipole transitions. Our results indicate the possibility to open up new interdisciplinary fields between superconductivity and optoelectronics.

\end{abstract}
\maketitle

Recent discoveries of new superconductors~\cite{Nagamatsu01,Kamihara08} boosted up the research fields with new experimental as well as theoretical possibilities. From a scientific viewpoint one great advantage of superconductivity is its long coherence time, and this is the most important feature for quantum information processing~\cite{Nielsen00}. Based on this fact superconducting qubits and their operations have been extensively studied via interactions of Cooper pairs with photons in the micro-wave band~\cite{10.1103/PhysRevLett.96.127006,Mooij99,Niskanen07,Clarke08,McDermott05}. During these operations with photon energies smaller than the energy gap of superconductivity (on the order of meV), the Cooper pairs are preserved. On the other hand, when photon energies become larger than the superconductivity gap, there arises a question how the Cooper pairs are involved in these optical interactions. For example, absorption of higher-energy photons (on the order of $\sim$eV) only results in the destruction of Cooper pairs, but it has been applied to promising high-speed single-photon detectors~\cite{Sobolewski03}. It is still unexplored what will take place with the counter process of photon emission from Cooper-pair states in this higher photon energy range.

In this letter, we demonstrate that electron Cooper pairs injected into a semiconductor can be highly involved in the interband transition and accelerate the photon generation processes. Electron Cooper pairs injected into a light emitting diode (LED) active layer by the proximity effect~\cite{Gennes99} accelerate the radiative recombination speed drastically below superconducting critical temperature under the high internal quantum efficiency of 100$\%$. Our new finding is the experimental demonstration of theoretically predicted Cooper-pair's gigantic oscillator strength~\cite{Hanamura02}, and this is well accounted for by a theoretical model considering the dipole transitions of condensed electron Cooper pairs with holes in the semiconductor valence band. 
\begin{figure}[h]
\includegraphics[width=18pc]{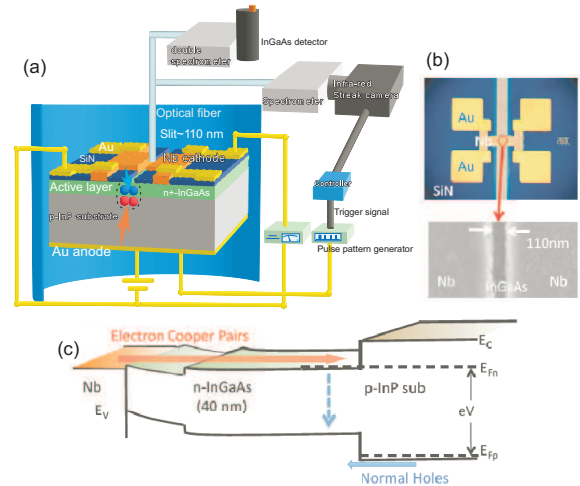}\hspace{2pc}%
\caption{\label{fig1} (Color online) Schematic of Cooper-pair LED and measurement setup: (a) Schematic of the LED structure and measurement setup of the I-V and EL properties. (b) Optical-microscope picture of the LED surface and the expanded secondary-electron-microscope picture of the slit formed in the Nb electrode. (c) Schematic of electron Cooper-pair injection from the surface Nb electrodes to the p-n junction, which recombine with holes in the valence band injected with the higher forward bias of the p-electrode. }
\end{figure}

Radiative recombination between electron Cooper pairs and hole Cooper pairs at a p-n junction of electron- and hole-superconductors was treated by one of the authors (E. H.) and coherent spontaneous emission, i.e., superradiance process was predicted theoretically~\cite{Hanamura02}. This concept was extended to the case of a semiconductor quantum dot (QD) and a method to generate regulated entangled photon pairs was proposed by the radiative recombination between electron Cooper pairs and a hole pair in a discrete QD energy level~\cite{Suemune02}. Toward this experimental verification, radiative recombination of electron Cooper pairs and holes in a single-heterostructure semiconductor is examined in this work.

The Cooper-pair injection into a semiconductor was studied with a light emitting diode (LED) in the optical-fiber communication band (Fig.~\ref{fig1}(a)). The LED epitaxial layers were grown on a p-type (001) InP substrate by metalorganic vapor-phase epitaxy. The layers consist of a 500-nm-thick p$^+$-InP buffer layer (Zn-doping $\sim$$1\times 10^{17}$ cm$^{-3}$), a 30-nm-thick n$^+$-In$_{0.53}$Ga$_{0.47}$As active layer (Si doping $\sim$$5\times 10^{18}$ cm$^{-3}$) lattice matched to InP, and a 10-nm-thick n$^{+}$-In$_{0.7}$Ga$_{0.3}$As ohmic contact layer (Si doping $\sim$$5\times 10^{18}$ cm$^{-3}$). Electron Cooper pairs are injected from the niobium (Nb) superconducting (SC) electrode into the conduction band of the n-InGaAs layers by the proximity effect~\cite{Gennes99}. The Nb electrode is 20-$\mu$m wide and 80-nm thick and is split by the 110-nm-wide slit formed by reactive ion etching (Fig.~\ref{fig1}(b)). The p-electrode was formed with a Au/Cr metal layers and was used as the anode for the LED operation as well as the back gate for the Nb/n-InGaAs/Nb I-V measurements. When the LED is forward biased across the p-n junction with the common voltage to the split Nb electrodes, electron Cooper pairs are injected into the 40-nm-deep n-InGaAs/p-InP junction through the conduction band of the n-InGaAs layers (Fig.~\ref{fig1}(c)) and recombine with holes in the valence band injected from the counter p-type electrode through the p-InP substrate. The LED was set in a $^3$He closed-cycle cryostat equipped with optical fibers. Electroluminescence (EL) is observed from the slit opening formed in the Nb electrode and was directly collected by a multi-mode optical fiber with the core diameter of 200 $\mu$m located $\sim$0.5 mm above the slit (Fig.~\ref{fig1}(a)). 

\begin{figure}[h]
\includegraphics[width=18pc]{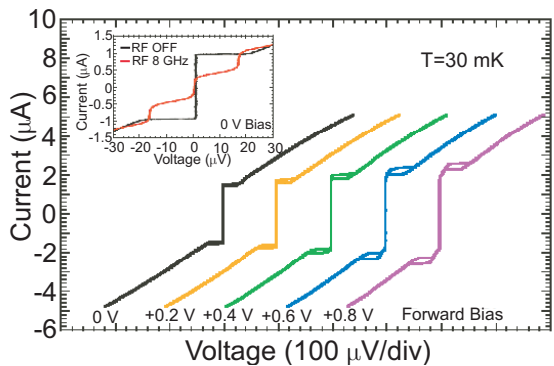}\hspace{2pc}%
\caption{\label{fig2} (Color online) Current-voltage characteristics of the Nb/n-InGaAs/Nb junction on the LED surface measured at 30 mK. The inset data measured at 8 GHz is shown but the Shapiro step intervals change with $\Delta V=2.1$ $\mu$V/GHz by changing the RF frequency. The I-V characteristics with the different forward bias are each shifted by 100 $\mu$eV for the display purpose. 
 }
\end{figure}

The Cooper-pair injection into the InGaAs semiconductor is examined by measuring the current-voltage (I-V) characteristics between the two niobium (Nb) superconducting electrodes on the LED surface (Fig.~\ref{fig1}(b)). The I-V characteristics show typical features of Josephson junctions (critical supercurrent of $I_{\rm C}$$\sim$1.2 $\mu$A) and the Shapiro steps~\cite{Duzer98} under the illumination of a radio-frequency (RF) signal (inset of Fig.~\ref{fig2}).
 That is, the two Nb electrodes together with n-InGaAs between them construct a superconductor/semiconductor/superconductor (S/Sm/S) junction, and Cooper pairs carry supercurrent across the SSmS junction. Moreover, we found that the Josephson junction characteristics are dependent on the forward bias applied to the p-n junction (Fig.~\ref{fig2}). When the current flowing through the p-n junction is negligible, the critical supercurrent $I_{\rm C}$ increases with the forward bias. The increase of the critical supercurrent is attributed to the increase of the n-InGaAs channel width in the Josephson junction (Fig.~\ref{fig1}(a)), which is modulated by the thickness of the depletion layer in the n-InGaAs side of the p-n junction. Therefore, we conclude that the superconducting Cooper pairs which are injected from the two Nb electrodes reach the light-emitting layer at the interface of the p-n junction.

\begin{figure}[h]
\includegraphics[width=18pc]{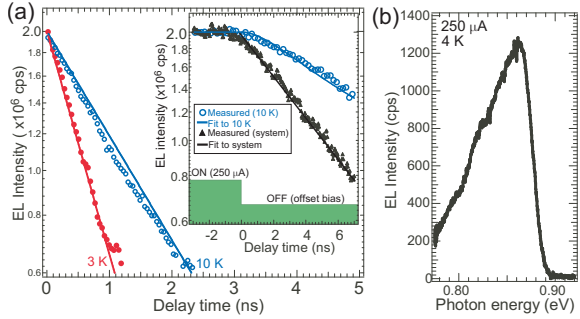}\hspace{2pc}%
\caption{\label{fig3} (Color online) 
(a) Transient decay of the spectrally integrated EL intensity: Diode bias set to the ON-state is stepwise changed to the OFF-state at 0 ns. Measured transient EL decay with the offset bias of zero and 600 mV shown with solid (black) triangles and open (blue) circles in the inset, respectively. The solid lines are the model fitting. The intrinsic EL time decays restored from the measured data are shown for 3 K and 10 K. (b) EL spectrum measured at 4 K with the constant injection current.
}
\end{figure}

The EL output is observed by forward biasing the p-n junction with the common voltage to the split Nb electrodes, and the typical EL spectrum measured at 4 K from the Nb slit of the LED is shown in Fig.~\ref{fig3}(b). The emission peak photon energy is 0.86 eV (the wavelength of 1.44 $\mu$m) and is in the optical-fiber communication band, which is determined by the energy gap of InGaAs lattice-matched to InP and n-type doping in the light emitting layer.

The impact of the electron-Cooper-pair injection on the LED operation appears in the radiative-recombination oscillator strength and is most efficiently visualized by the lifetime measurements. For this purpose the injection current was step-wise decreased from the steady-current ON-state to the offset-biased but negligible-current OFF-state as shown in the inset in Fig.~\ref{fig3}(a), and the resultant transient EL decay was examined. When the system capacitance-resistance (CR) time constant is included with the notation of $\tau _{\rm CR}$, the transient decay of the spectrally integrated EL intensity is given by, $I_{\rm EL}(t)=J_{0}\eta _{\rm int}\eta _{\rm det} \left \{ e^{-t/\tau _{\rm LED}} +(1-\tau _{\rm LED}/\tau _{\rm CR})^{-1} \left ( -e^{-t/\tau _{\rm LED}}+e ^{-t/\tau _{\rm CR}} \right ) \right \}$ for $t\ge 0$. $\tau _{\rm LED}$ is the EL decay time constant, $J_{0}$ is the injection current divided by the electric charge, $\eta _{\rm int}$ is the LED internal quantum efficiency, and $\eta _{\rm det}$ is the EL detection efficiency. The observed transient decay was steeper for the lower offset bias and approached to the single exponential decay with the zero or the reverse offset bias (solid (black) triangles in the inset of Fig.~\ref{fig3}(a)). This steeper decay is attributed to the internal-electric-field-induced Stark effect that spatially separates electrons and holes~\cite{PhysRevB.72.033316} or tunneling back of the injected holes to p-InP. Under this condition, $\tau _{\rm LED} \ll \tau _{\rm CR}$ and the above equation is simplified to $I_{\rm EL}(t)=J_{0}\eta _{\rm int}\eta _{\rm det}e^{-t/\tau _{\rm CR}}$. This determines $\tau _{\rm CR}$ to be 2.70 ns.

The diode current was set constant to 250 $\mu$A for the ON-state and the temperature dependence of the transient decay was measured by setting the offset bias to 600 mV for the OFF-state. The latter offset bias was selected to satisfy both the negligible diode current less than 1 nA and the negligible internal-field effect based on a series of measurements on the bias dependence. The ON-state EL intensity remained almost constant against the temperature. However the decay time showed the clear temperature dependence. One of the data measured at 10 K is shown in the inset of Fig.~\ref{fig3}(a) (open (blue) circles). From the fit with the solid line considering the CR time constant, the lifetime $\tau _{\rm LED}$ of 2.27 ns was derived. Once the time constants are fixed in this way, the intrinsic EL time decay can be restored from the measured EL intensities and is reproduced in Fig.~\ref{fig3}(a) for 3 K and 10 K. The reduction of the lifetime for the lower temperature is clearly observed.

\begin{figure}[h]
\includegraphics[width=18pc]{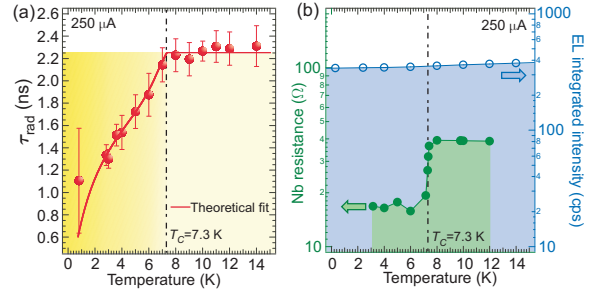}\hspace{2pc}%
\caption{\label{fig4} (Color online) 
(a) Temperature dependence of the measured LED radiative recombination lifetime, $\tau _{\rm rad}$, and theoretical examination (The larger error bar at the point at 0.8 K is due to the limited measurement time coming from the heating of the shunt resistance.) The solid (red) line is the fit with the theoretical model based on the Cooper-pair radiative recombination. (b) Resistance (solid (green) circles) measured along the Nb electrode. The integrated EL intensity (open (blue) circles) during the ON-state is almost temperature independent. 
}
\end{figure}

The temperature dependence of the measured lifetime $\tau _{\rm LED}$ is summarized in Fig.~\ref{fig4}. It is almost constant at $\sim$2.25 ns above the temperature of $\sim$7.5 K, but then the abrupt shortening below that temperature is clear. It is shortened up to 1.1 ns at 0.8 K. The resistivity of the Nb electrode was measured employing the two neighboring Au pads (Fig.~\ref{fig1}(b)) and is shown in Fig.~\ref{fig4}(b). It is abruptly reduced at the temperature of 7.3 K and this shows the superconducting critical temperature, $T_{\rm C}$, of the Nb-electrode in this LED sample~\cite{Add2}. The formation of electron Cooper pairs initiates in the Nb electrode below $T_{\rm C}$ and the onset temperature of the EL lifetime shortening precisely agrees with $T_{\rm C}$. This agreement as well as the abrupt lifetime shortening below $T_{\rm C}$ demonstrates the major role of the injected Cooper pairs for the accelerated recombination rates in the LED. The final important experimental step is to determine the main recombination mechanism which dominates the observed EL lifetime. In spite of the significant change of the measured lifetime, the integrated EL intensities remained almost constant against the temperature even across the superconducting critical temperature (Fig.~\ref{fig4}(b)). This is completely different from the previous observation of the drastic EL enhancement below $T_{\rm C}$~\cite{Hayashi08}. This is due to the difference of the internal quantum efficiencies as discussed below. The temperature dependence of the EL intensity is determined by the LED internal quantum efficiency $\eta _{\rm int}$, which is given by the balance of the radiative and nonradiative recombination rates as follows: $\eta _{\rm int}=1/\tau _{\rm rad}/(1/\tau _{\rm rad}+1/\tau _{\rm nonrad})$, where $\tau _{\rm rad}$ and $\tau _{\rm nonrad}$ are the radiative and nonradiative lifetimes, respectively. The measured EL lifetime is given by  $\tau _{\rm LED}=1/(1/\tau _{\rm rad}+1/\tau _{\rm nonrad})$. The condition to satisfy the present situation of the constant $\eta _{\rm int}$ in spite of the shortening of $\tau _{\rm LED}$ below $T_{\rm C}$ is uniquely determined so that $\tau _{\rm LED}\approx \tau _{\rm rad} \ll \tau _{\rm nonrad}$ and $\eta _{\rm int}\approx 1$. This naturally leads to the $\sim$100 \% internal quantum efficiency of the LED, and the abrupt lifetime shortening below $T_{\rm C}$ shown in Fig.~\ref{fig4}(a) is that of the radiative lifetime $\tau _{\rm rad}$.

The observed radiative recombination enhancement is examined with a theoretical model considering Cooper pairs~\cite{Hanamura02,Suemune02}. Based on the theoretical analysis of the proximity effect at the S-Sm interface by P. G. de Gennes~\cite{Gennes99}, the Cooper-pair amplitude is given by $F={\cal h}\varphi _{\uparrow}(\vec {r})\varphi _{\downarrow}(\vec {r}){\cal i}$, where $\varphi _{\uparrow}(\vec {r})\varphi _{\downarrow}(\vec {r})$ is the spin-paired field operator and ${\cal hi}$ is the ensemble average. In a uniform superconductor, $F$ is proportional to the pair potential $\Delta _{0}$~\cite{Gennes99}, which is equal to the superconductivity gap. The penetration of the pair amplitude from the S-Sm interface into the n-InGaAs region was treated in the dirty limit and was shown to have the spatial dependence of $e^{-z/\xi _{\rm N}}$~\cite{Parks69,Ketterson99}. $\xi _{\rm N}$ is the coherent length in the normal region and the distance $z$ is measured from the S-Sm interface toward the p-n junction (Fig.~\ref{fig1}(c)). Based on this understanding the electric-dipole transition~\cite{Hanamura02} between the electron Cooper pairs injected into the n-InGaAs conduction band and holes injected in the valence band was studied and the formula is derived for the radiative recombination rate in the following intuitively understandable form: 
\begin {eqnarray}
W_{\rm super}&=&A\frac {\Delta _{0}^{2}(T)}{T}\exp (-2L/\xi _{\rm N}(T)).
\label {eq2}
\end {eqnarray}
$A$ is the constant that includes the dipole moment. $\Delta _{0}^{2}(T)$ is the square of the temperature-dependent pair potential and is proportional to the Cooper-pair number in the superconductor. This factor appears because the number of the electron Cooper pairs penetrated into the recombination region is proportional to this factor in the superconductor. The exponential factor originates from the pair-amplitude penetration into the InGaAs layer. The coherent length $\xi _{\rm N}$ of the electron Cooper pairs penetrated into the n-InGaAs layer is given in the dirty limit by the following equation~\cite{Akazaki89}, $\xi _{\rm N}(T)=\left ( \hbar ^{3}\mu/2\pi k _{\rm B} T m_{\rm e} {\rm e} \right )^{1/2} \left ( 2\pi ^{2}N_{\rm 3D} \right ) ^{1/3}$, where $\hbar $ and $k_{\rm B}$ are the Planck constant and the Boltzman constant. For the In$_{0.53}$Ga$_{0.47}$As layer, the electron mobility was assumed to be $\mu \sim$4000 cm$^{2}$/Vs based on measurements at 77 K and room temperature. For the electron concentration of $N_{\rm 3D}\sim$$5\times 10^{18}$ cm$^{-3}$ and the electron effective mass of $m_{\rm e}/m_{0}=0.043$~\cite{Vurgaftman01}, the coherent length is calculated to be $1570/\sqrt {T}$ nm. This gives 1570 nm at 1 K and 785 nm at 4 K. The distance from the S-Sm interface to the recombination region is given as $L$. In the present LED, $L$ is equal to 40 nm and therefore is much shorter than $\xi _{\rm N}$. The factor $T$ in the denominator originates from the volume integral to calculate the transition probability. Since the active region of the present LED is limited by the InGaAs thin layer, the volume integral is limited to the coherent volume of the Cooper pairs, that is, on the order of $L\xi _{\rm N}^{2}$. This results in the factor $T$ in the denominator. The formula given by Eq.~(\ref{eq2}) was also derived separately with the second-order perturbation theory on the radiative recombination of an electron Cooper pair and two p-type carriers (holes)~\cite{Asano09}.

The radiative lifetime that includes the role of Cooper pairs is given by $\tau _{\rm rad}=[ (1/\tau _{\rm rad})_{\rm Normal}+W_{\rm Super} ] ^{-1}$. The radiative lifetime measured at the temperature above $T_{\rm C}$ is used as the first term in this equation. For the comparison with measurements, the constant A in Eq.~(\ref{eq2}), the relative weight of the Cooper-pair contribution, is the just one fitting constant. It is noted that the essential character, the abrupt decrease of the radiative lifetime below $T_{\rm C}$, is well accounted for by the present model (solid (red) line in Fig.~\ref{fig4}(a)).

In the present radiative recombination of conduction-band electrons in the superconducting state with the holes in the normal state, the temporal coherency will be lost because of the energy distribution of the holes in the valence-band. However, the spatial coherency of the electron-Cooper pairs enhances the radiative decay rate because of the giant oscillator strength~\cite{Hanamura73}. Here a bound state of the Cooper pairs extends over the coherent length so that any holes within the coherent volume can radiatively decay with one of the Cooper pair with the same spin orientation. This effect is observed as the present enhancement of the radiative recombination rate. Therefore the essential physics in the present scheme relies on the fact that Cooper pairs are the condensation of electrons at a spin-singlet quantum state. The dominant factor determining the temperature dependence in Eq.~(\ref{eq2}) is the pair potential term $\Delta _{0}^{2}(T)$ [\onlinecite{Duzer98,Schrieffer}], and this term is proportional to the Cooper-pair number in the superconducting state. The higher condensation at the lower temperature shows up the Cooper-pair's gigantic oscillator strength more clearly in our lifetime measurements. The recombination of electron Cooper pairs near the quasi-Fermi level in the conduction band with holes in the valence band may be reflected in the measured EL spectra. At present such a correspondence is not very clear. The reason will be the broadening of the hole energy distribution but the details are still under study. Yet this new finding of the Cooper-pair's gigantic oscillator strength in semiconductor interband optical transitions opens up new interdisciplinary fields between superconductivity and optoelectronics.

We thank M. Jo for potential profile calculations of the LED structure, H. Kan and M. Yamanishi for valuable discussion on our measurements. Supports from the Hokkaido Innovation through Nanotechnology Support, Post-Silicon Materials and Devices Research Alliance, and Global COE, Graduate School of IST, Hokkaido University are acknowledged.

\bibliography{sqled4}

\begin{thebibliography}{10}
\bibitem{Add2}%
  \BibitemOpen
  \bibinfo {note} {The measured Nb $T_{\rm C}$ is slightly lower than the
  intrinsic value of 9.3 K. This is due to unintentional evaporation conditions
  of the Nb metal. The residual resistance below $T_{\rm C}$ is attributed to
  thin oxide layers at the Au-Cr pad/Nb interface formed during the patterning
  processes, but this does not influence the proximity effect at the
  Nb/n-InGaAs superconductor-normal (S-N) interface.}%
  \bibAnnoteFile{Stop}{Add2}%
\end{thebibliography}

\end{document}